# HT-MMIOW: A Hypothesis Test approach for Microbiome Mediation using Inverse Odds Weighting

Yuka Moroishi[1,2], Zhigang Li[3], Juliette C. Madan[2,4], Hongzhe Li[5], Margaret R. Karagas[2], Jiang Gui[1*]

Author Affiliation

[1]Department of Biomedical Data Science, Geisel School of Medicine at Dartmouth, Hanover, New Hampshire 03755, USA

[2]Department of Epidemiology, Geisel School of Medicine at Dartmouth, Hanover, New Hampshire 03755, USA

[3]Department of Biostatistics, University of Florida, Gainesville, Florida 32603, USA

[4]Department of Pediatrics, Children's Hospital at Dartmouth, Lebanon, New Hampshire 03766, USA

[5]Department of Biostatistics and Epidemiology, University of Pennsylvania Perelman School of Medicine, Philadelphia, Pennsylvania 19104-6021, USA

* To whom correspondence should be addressed: Jiang Gui <Jiang.Gui@dartmouth.edu>.

Abstract

The human microbiome has an important role in determining health. Mediation analyses quantify the contribution of the microbiome in the causal path between exposure and disease; however, current mediation models cannot fully capture the high dimensional, correlated, and compositional nature of microbiome data and do not typically accommodate dichotomous outcomes. We propose a novel approach that uses inverse odds weighting to test for the mediating effect of the microbiome. We use simulation to demonstrate that our approach gains power for high dimensional mediators, and it is agnostic to the effect of interactions between the exposure and mediators. Our application to infant gut microbiome data from the New Hampshire Birth Cohort Study revealed a mediating effect of 6-week infant gut microbiome on the relationship between maternal prenatal antibiotic use during pregnancy and incidence of childhood allergy by 5 years of age.

# 1. Introduction

The human gut microbiome and immune system interact to form a bidirectional relationship (Zheng *et al.*, 2020), and the developing gut microbiome plays a crucial role in immune system maturation in infancy (Gensollen *et al.*, 2016). As a result, perturbations in the gut microbiome is linked to clinical outcomes through infancy and childhood, and into adulthood. External factors shape the gut microbiome including delivery mode, diet, and antibiotic use (Hasan and Yang, 2019). Understanding the three-way interplay among external factors, the gut microbiome, and clinical outcomes can help generate opportunities for intervention such as therapeutics and probiotics for modulating the gut microbiome to improve health outcomes.

Mediation analysis quantifies the contribution of a variable in the causal pathway between an exposure or treatment and an outcome. These analyses are especially useful in health research to determine possible interventions to prevent disease or produce better health outcomes. Upon development of approaches to mediation analysis including traditional approaches (Baron and Kenny, 1986; MacKinnon *et al.*, 2002) and causal inference approaches (Robins and Greenland, 1992; Pearl, 2001; Rubin, 2005; VanderWeele and Vansteelandt, 2010; Imai and Yamamoto, 2013), many extensions of mediation analyses have also been published in recent years. These include models for multiple mediators (Imai and Yamamoto, 2013; VanderWeele and Vansteelandt, 2014) and those that account for interactions between exposures and mediators (Valeri and Vanderweele, 2013; VanderWeele, 2014). Other approaches include inverse odds ratio weighting (IORW) (Tchetgen Tchetgen, 2013) or inverse odds weighting (IOW) (Nguyen *et al.*, 2015) to quantify the total, direct, and indirect effects. Testing the mediation effect can be conducted using common approaches such as the Sobel test (Sobel, 1982) and the joint significance test (MacKinnon *et al.*, 2002).

Mediation analysis for high dimensional mediators has become increasingly popular for modeling biomarkers. These include methods using PCA and regularization for dimension reduction (Huang and Pan, 2016; Chén *et al.*, 2018; Zhang, 2021; Zhao *et al.*, 2020), multiple testing procedures (Boca *et al.*, 2014; Millstein *et al.*, 2016; Sampson *et al.*, 2018; Djordjilović *et al.*, 2019; Liu *et al.*, 2022; Dai *et al.*, 2022), and others (Gao *et al.*, 2019; Song *et al.*, 2020). Methods for mediation analysis have also extended to microbiome data: many test for mediation with continuous outcome (Sohn and Li, 2019; Zhang *et al.*, 2018; Wang *et al.*, 2020; Zhang, Chen, Feng, *et al.*, 2021; Zhang, Chen, Li, *et al.*, 2021; Wu *et al.*, 2021), some account for interactions between exposure and mediators (Wang *et al.*, 2020; Wu *et al.*, 2021), and one conducts mediation analysis for dichotomous outcomes (Sohn *et al.*, 2022). Most of these methods, however, require specification of the microbiome model, which is difficult due to the sparse, high dimensional, and compositional nature of microbiome data. Furthermore, to our knowledge, none of them can accommodate both dichotomous outcomes and high dimensional microbiome mediators.

We propose a novel approach, HT-MMIOW, a Hypothesis Test approach for Microbiome Mediation using IOW, that tests the mediation effect of high dimensional microbiome data on both continuous and dichotomous outcomes using IOW. This approach uses the isometric log-ratio transformation (ilr) to account for the compositional nature of microbiome data and reduces dimensions using the Uniform Manifold Approximation and Projection (UMAP). The components generated from UMAP serve as mediator variables. A permutation test is performed to determine the statistical significance of the overall microbial mediation effect. Our simulation results demonstrate that this new approach is well powered when the number of true mediators is large or the indirect effect is large. We present an application of our approach to infant gut microbiome data to detect a mediating effect between antibiotic exposure and a diagnosis of allergy before the age of 5.

## 2. Methods

### 2.1 Identification

Suppose the exposure, mediator, and outcome are denoted by an $n \times 1$ vector $E$, an $n \times p$ matrix $M$, and an $n \times 1$ vector $Y$ respectively. $Y$ is known to succeed $M$, and $M$ is known to succeed $E$. The microbiome matrix $M$ is composed of abundance of $p$ taxa for each subject in $n$, and there are $t$ true mediators in $M$. For simplicity, let both $E$ and $Y$ be dichotomous. Let $X$ be an $n \times q$ matrix of covariates. Under the Baron and Kenny's classical mediation framework for a single mediator $M$, mediation analysis is represented by the following equations:

$$logit(\Pr(Y = 1 \mid e)) = a_0 + a_1 E$$

$$E[M|e] = b_0 + b_1 E$$

$$logit(\Pr(Y = 1 \mid e, m)) = c_0 + c_1 E + c_2 M$$

The direct effect of $E$ on $Y$ is $c_1$, the indirect effect of $E$ on $Y$ through $M$ is $b_1 c_2$, and the total effect is $a_1 = c_1 + b_1 c_2$. The Sobel test determines the existence of an indirect effect with the hypothesis $H_0: b_1 c_2 = 0$ vs $H_A: b_1 c_2 \neq 0$.

## 2.2 Dimension reduction of microbiome data

The standard mediation model requires the knowledge of true mediators. Mediation models for multiple mediators exist, but many assume that mediators are independent or that the causal relationship between mediators are known. However, microbiome data is high dimensional, compositional, and correlated. We do not know which microbial taxa are true mediators, and we do not know the causal relationships among taxa. We propose mapping the microbiome data from the Aitchison-simplex to Euclidean space and reducing the dimensions to obtain a smaller set of mediation features.

The ilr is an extension of log-ratio transformations for compositional data that extends on traditional additive (alr) and centered (clr) log-ratio transformations. Briefly, alr is the log-ratio of a value in the composition and a reference value, and clr is the log-ratio of a value and the geometric mean of values in the composition. There are, however, limitations to these approaches; alr produces transformations do not preserve distances, and clr produces transformations with a singular covariance matrix. In ilr, distance is preserved when data is transformed from the $p$-dimensional Aitchison space to the $p-1$ dimensional Euclidean space, and the resulting vectors are orthogonal and interpretable in analyses (Egozcue *et al.*, 2003).

Each value of ilr transformation is a "balance" between two subsets of $M$, denoted by $R$ for those on the left of the balance and $S$ for those on the right of the balance:

$$ilr(R, S) = \sqrt{\frac{rs}{r+s}} \log\left(\frac{g(m_R)}{g(m_S)}\right)$$

where $r$ is the cardinality of $R$, $s$ is the cardinality of $S$, $m_R$ are the values in $R$, $m_S$ are the values in $S$, and $g(\cdot)$ is the geometric mean function.

We impute a pseudocount of 0.5 to zero values and apply ilr to the microbiome data to account for compositionality. After ilr, we reduce dimensionality using UMAP. Briefly, UMAP is a fast dimension reduction procedure that models the manifold with a fuzzy topological structure by searching for a low-dimensional projection with the closest possible equivalent fuzzy topological structure (McInnes *et al.*, 2020). UMAP allows for non-linear dimension reduction and meaningful separation between clusters. Applying UMAP on $ilr(M)$ transforms our microbiome data to an $n \times c$ matrix $U$, where $c$ is a user-specified number of components.

## 2.3 Compute the indirect effect and perform permutation test for mediation

The IORW and IOW approach for causal mediation analysis allows for the decomposition of the total effect to direct effect and indirect effect. This approach can accommodate multiple mediators due to a weighting procedure that removes the need to specify the model to regress the multiple mediators on exposure. This approach is advantageous for microbiome data because of the difficulty in modeling the joint conditional density of high dimensional, compositional, and correlated microbiome features. Using the IOW approach, the total effect of exposure $E$ on outcome $Y$, given covariates $X$, can be estimated using the model:

$$h(Y|E, X) = \beta_0 + E\beta_1 + X\beta_2 \quad (1)$$

where $\beta_1$ represents the total effect of $E$ on $Y$, given $X$. $h(\cdot)$ is the user-specified link function and $\varepsilon$ represents the error. We implement IOW to condense the association between $E$ and mediators $U$, conditional on $X$. The weights can be estimated using the model with a user-specified link function $j(\cdot)$:

$$j(E|U, X) = \alpha_0 + U\alpha_1 + X\alpha_2$$

$$(2)$$

For each observation, the weight is the inverse of the predicted odds in the exposed group and 1 for the unexposed group. Using IOW as opposed to IORW stabilizes the weights, though it may introduce small bias to estimates (Nguyen *et al.*, 2015). We then use a weighted regression model to estimate the direct effect of $E$ on $Y$, conditional on .

$$h(WY|WE, WX) = \gamma_0 W + WE\gamma_1 + WX\gamma_2$$

$$(3)$$

where $\gamma_1$ represents the direct effect of $E$ on $Y$, given $X$, and $W$ is an $n \times 1$ vector of weights. Due to this weighting procedure, we do not need to specify interactions between $E$ and $U$.

We can now estimate the indirect effect using parameters from (1) and (3), which becomes our observed test statistic for our permutation test: $T_{obs} = \beta_1 - \gamma_1$. The null hypothesis of no mediation effect of the microbiome is expressed by:

$$H_0 : \beta_1 - \gamma_1 = 0,$$

and the alternative hypothesis that a mediation effect of the microbiome exists is expressed by:

$$H_A : \beta_1 - \gamma_1 \neq 0.$$

We apply permutation test to calculate the P-value using the formula $\left(\sum_{j=1}^{B} I(|T^{(i)}| > |T_{obs}|)\right)/B$, where B is the total number of permutations, $I(\cdot)$ is an indicator function, $T^{(j)}$ is the test statistic under the null hypothesis for permutation $j = 1, \ldots, B$.

## 3. Simulation

### 3.1 Simulation overview

We conducted simulations to evaluate the power of our proposed approach on continuous and dichotomous outcomes. We simulated $E$ and $\boldsymbol{M}$ using SparseDOSSA (Ma *et al.*, 2021). Briefly, this tool uses zero-inflated log-normal distribution to simulate realistic microbial community structure of human stool and covariates that correlate with the simulated microbiome data. We applied UMAP on $\boldsymbol{M}$ to reduce dimensionality to 2 components. We set the percentage of microbial features associated with the exposure at 50% and effect size of the exposure on mediator at 3 for simulated data generation using SparseDOSSA. We simulated outcomes $Y$ using a standard logistic regression with the exposure and scaled relative abundance of true microbial taxa mediators selected at random from those associated with the exposure, with the effect size of the exposure on outcome set at 5 and a random noise of of $N(0,1)$. Exposure variables were dichotomous, and outcome variables were continuous and dichotomous. We also evaluated the type I error under the null hypothesis with varying sample sizes ($n$ = 50, 70, 100, 150, 300, 500) and number of taxa ($p$ = n, 2n) for continuous and dichotomous outcomes.

To fine-tune our approach, we evaluated the performance of HT-MMIOW with various dimension reduction techniques. These include 1) UMAP as described previously; 2) PCA to n-components then UMAP; 3) PCA with components explaining 100% of the variance; and 4) PCA with components explaining 80% of the variance. We investigated varying effect sizes of true mediators on outcome (effect size = 0.5, 1, 5), which serve as proxy for smaller to larger indirect effects, and varying number of true mediators in the microbiome sample ($t$ = 1, 5, 10). We also varied sample sizes ($n$ = 50, 70, 100, 150, 300, 500) and the number of taxa in the microbiome data ($p$ = n, 2n). We compared the performance of HT-MMIOW with a distance-based omnibus test using Bray-Curtis and Jaccard distances (Zhang *et al.*, 2018). Though the omnibus test is designed for continuous outcomes, we sought to examine its performance on dichotomous outcomes due to the lack of methods for high dimensional microbiome mediators and dichotomous outcomes. All analyses were conducted using R version 4.0.2 and packages "SparseDOSSA2", "magrittr", "dplyr", "ggplot2", "compositions", "umap", "foreach", "doParallel", and "bda".

## 3.2 Simulation results

Figures 1 and 2 display the simulation results of HT-MMIOW for continuous outcomes under varying conditions compared with the omnibus test. HT-MMIOW with UMAP dimension reduction performs better than HT-MMIOW with PCA and UMAP when the number of taxa are twice the sample size. HT-MMIOW also performs better than the distance-based omnibus test for all conditions, especially when the number of true mediators is small. Using HT-MMIOW with only PCA as the dimension reduction technique did not perform well. HT-MMIOW with PCA components explaining 80% of the variance generally performed better than HT-MMIOW with PCA components explaining 100% of the variance, and this may be because fewer components in the inverse odds ratio mediation approach provides more power. These power calculations were based on an empirical threshold of 0.05. Though type I error for continuous outcomes were around 0.05 based on the 0.05 empirical threshold, type I error based on an empirical threshold of 0.01 were smaller (Supplementary Figure 1). HT-MMIOW using a 0.01 empirical threshold continued to yield higher power than the omnibus distance test (Supplementary Figure 2).

For dichotomous outcomes, HT-MMIOW performed well with stronger effects of mediators on Y and with larger numbers of true mediators in M (Figure 3, Figure 4). Using UMAP performed just as well as using PCA and UMAP. HT-MMIOW also performed better than the distance-based omnibus test in all conditions. Again, HT-MMIOW's performance was lower when using only PCA as the dimension reduction technique. As expected, performances for both continuous and dichotomous outcomes increased with with increasing sample size. Again, the above power calculations were based on an empirical threshold of 0.05. The type I error for dichotomous outcomes were inflated when an empirical threshold of 0.05 was used; however, type I error based on an empirical threshold of 0.01 were close to zero (Supplementary Figure 1). HT-MMIOW using a 0.01 empirical threshold for dichotomous outcomes yielded higher power than the omnibus distance test (Supplementary Figure 2).

## 4. Application to NHBCS Data

Prenatal antibiotic use has been linked to development of infant and childhood allergy (Baron *et al.*, 2020), and it is also associated with compositional differences in the infant gut microbiome (Dierikx *et al.*, 2020; Coker *et al.*, 2020). We applied HT-MMIOW to test the effect of the infant gut microbiome as a potential mediator in the causal path between prenatal antibiotic use and allergy on data from the New Hampshire Birth Cohort Study (NHBCS). The NHBCS is a prospective birth cohort of mother-infant dyads who received prenatal care in New Hampshire clinics. The study recruited pregnant women with ages ranging between 18 and 45, who had a singleton pregnancy and were served by a private water system as described previously (Gilbert-Diamond *et al.*, 2011). Prenatal use of antibiotics was reported in prenatal records. The child's caregivers reported whether the child had any allergies diagnosed by a physician during telephone interviews conducted when infants turned approximately 4, 8, 12 and 18 months of age, and then at one year intervals thereafter. Infant stools samples were collected at approximately 6 weeks of age and sequenced using Illumina MiSeq (Illumina, San Diego, CA) for bacterial 16S rRNA gene sequencing of the V4-V5 hypervariable region at Marine Biological Laboratory in Woods Hole, Massachusetts. We inferred amplicon sequence variants using DADA2 (Callahan *et al.*, 2016) and assigned taxonomies using the SILVA database (Pruesse *et al.*, 2007).

A total of 412 mother-infant dyads were included in this analyses. Of these, 72 (17.5%) self-reported to antibiotic use during pregnancy, and 39 children (9.5%) had been diagnosed with allergy by 5 years of age. Based on the 16S sequencing reads of our 6-week stool samples, we identified 705 different genera of bacteria.  In our mediation test, HT-MMIOW produced a P-value of 0.008; this gives us evidence to reject the null hypothesis and suggest that the infant gut microbiome mediates the relationship between antibiotic use during pregnancy and incidence of allergy. Due to variable duration of follow-up, we adjusted for age at which allergy was first diagnosed or age at last follow-up in our model.

## 5. Discussion

We proposed a novel hypothesis test for microbiome mediation effect that utilizes a dimension reduction procedure for microbiome data and a mediation analysis procedure utilizing IOW to test if an indirect effect exists. Our simulation scenarios evaluated the power of our approach under varying conditions. HT-MMIOW was highly powered when the effect of the mediator on the microbiome was large or when the number of true mediators in the microbiome data was large. Compared to a conventional hypothesis test, HT-MMIOW generally performed better for both continuous and dichotomous outcomes. Our approach is one of the only hypothesis tests that can test for the indirect effect in high dimensional microbiome mediators and dichotomous outcomes.

Our proposed hypothesis test is flexible in its application. HT-MMIOW can be used for multiple types of exposure and outcome data; the user may specify their own link functions in the regression models. Our approach can also be adjusted to account for other types of high dimensional mediation analysis, including genomics, by replacing the centered log-ratio transformation step to a transformation of the user's choice. The total effect, direct effect, and indirect effect can be calculated using the IOW approach, and confidence intervals can be estimated using bootstrap methods.

The main strength of our approach is that HT-MMIOW reduces high-dimensional microbiome data to a few independent components that are representative of microbial community structure. The IOW procedure accommodates multiple mediators in the mediation model. Furthermore, the IOW frameworks eliminates the need to specify a model to regress the exposure on the joint conditional density of multiple compositional mediators. Another strength is that IOW does not require us to specify the interaction between exposure and mediators, even if they exist.

Despite its strengths, HT-MMIOW is not without limitations. One limitation of our approach is that the mediation effect of true microbial mediators must be large for the approach to detect an indirect effect. We must also assume that there is no unmeasured confounding. Additionally, the number of true mediators must be large for smaller mediation effects and smaller samples sizes. Our method also cannot detect the mediation effect of specific microbes or clusters of microbes. Nevertheless, this approach serves as an important first step in determining whether a mediation effect exists. Further work is warranted to identify key microbial features and interactions responsible for driving this effect.

Acknowledgements

We thank all participants and staff of the New Hampshire Birth Cohort Study. We also thank Hilary G. Morrison for processing the infant stool samples.

Funding

This work was supported by US National Institutes of Health under award numbers UH3OD023275, NIEHS P01ES022832, and NIGMS P20GM104416 and the US Environmental Protection Agency under award numbers RD83544201.

Author Contributions

Y.M., M.R.K., and J.G. conceived the study. Y.M. developed the statistical method, ran the simulations, and implemented the R package. Z.L., H.L., and J.G. contributed to the development of the method. M.R.K. and J.C.M. contributed to data acquisition and processing. Y.M. drafted the manuscript. All authors reviewed and edited the manuscript.

All authors declare no conflicts of interest.

Data and Code Availability

The microbiome data used in this study can be found under accession number PRJNA296814 at the online repository http://www.ncbi.nlm.nih.gov/sra. Code is available upon request. The R package for HT-MMIOW is available on GitHub: https://github.com/yukamoro/HTMMIOW

Figures

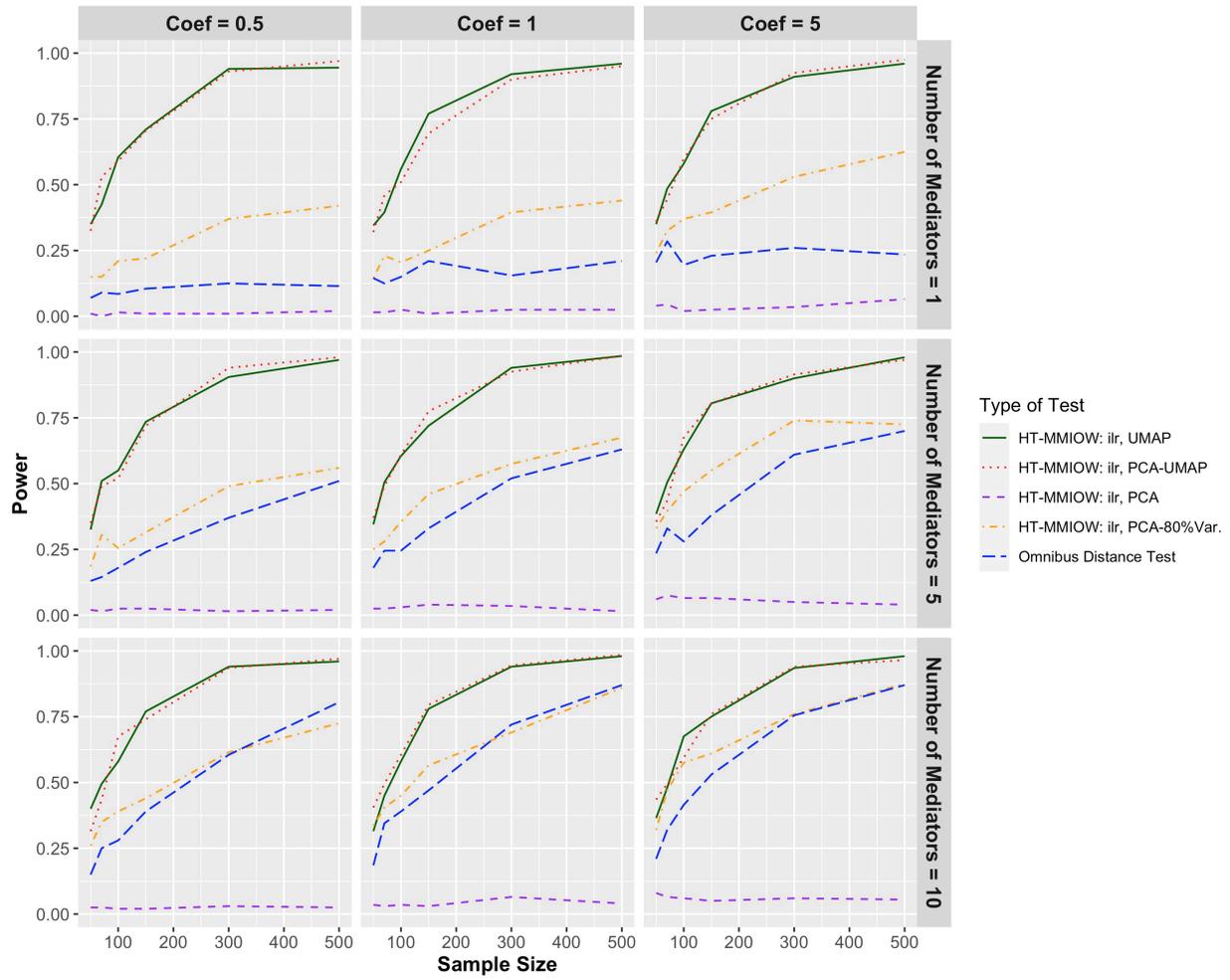

Figure 1. Power calculations for varying sample size, effect size, and number of true mediators for continuous outcomes and $p = n$. Results are based on 200 simulations. Each line represents the hypothesis test procedures evaluated.

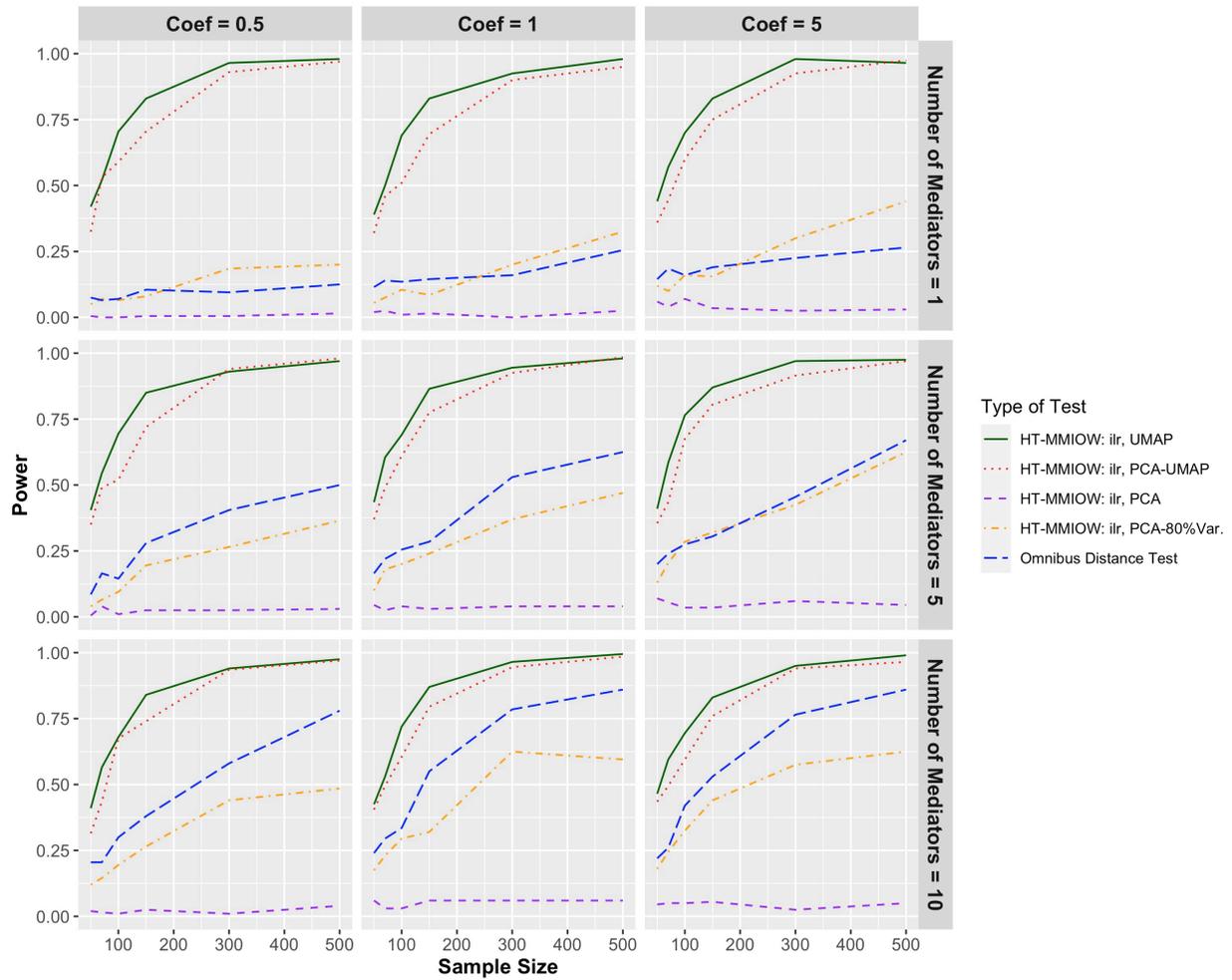

Figure 2. Power calculations for varying sample size, effect size, and number of true mediators for continuous outcomes and $p = 2n$. Results are based on 200 simulations. Each line represents the hypothesis test procedures evaluated.

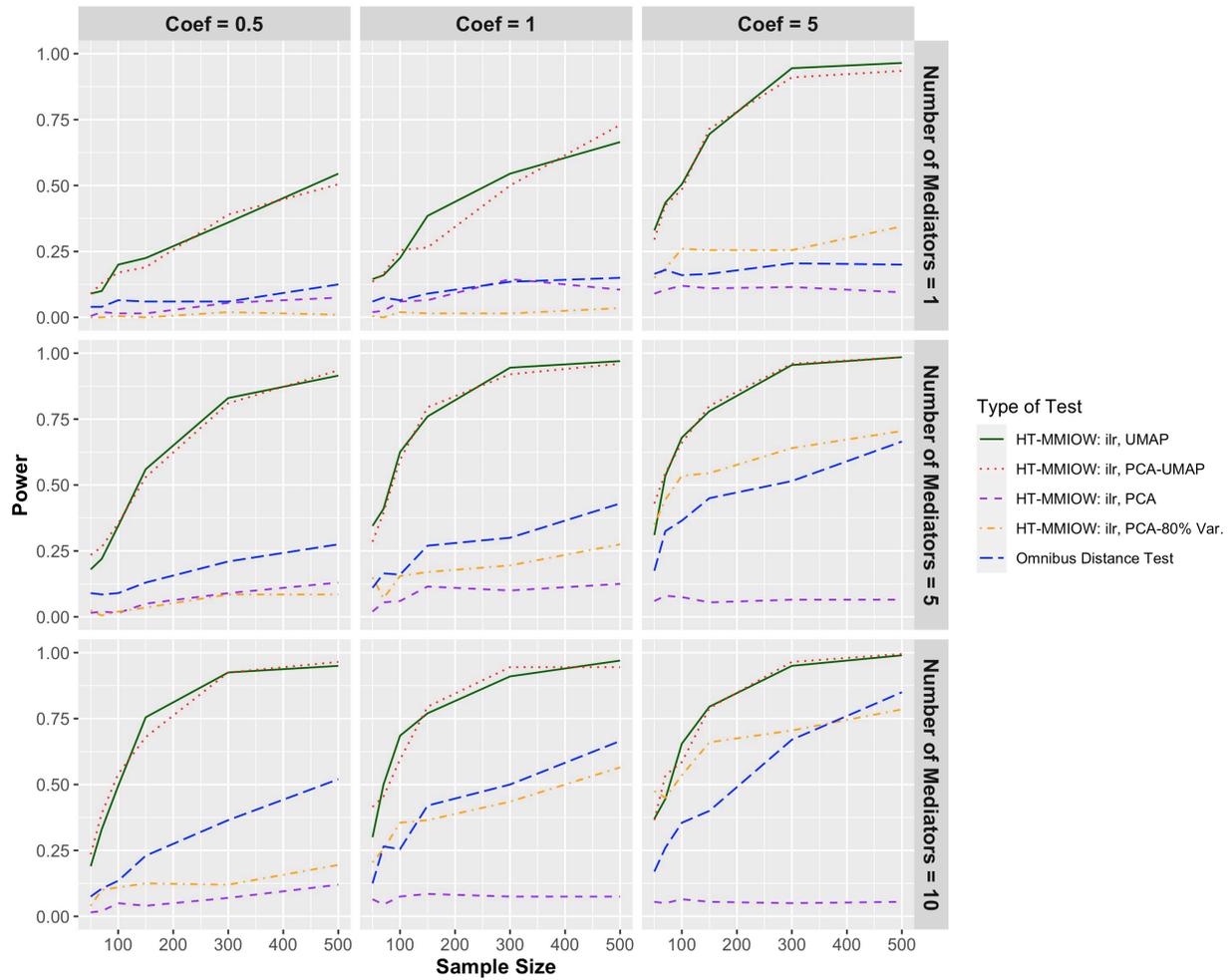

Figure 3. Power calculations for varying sample size, effect size, and number of true mediators for dichotomous outcomes and $p = n$. Results are based on 200 simulations. Each line represents the hypothesis test procedures evaluated.

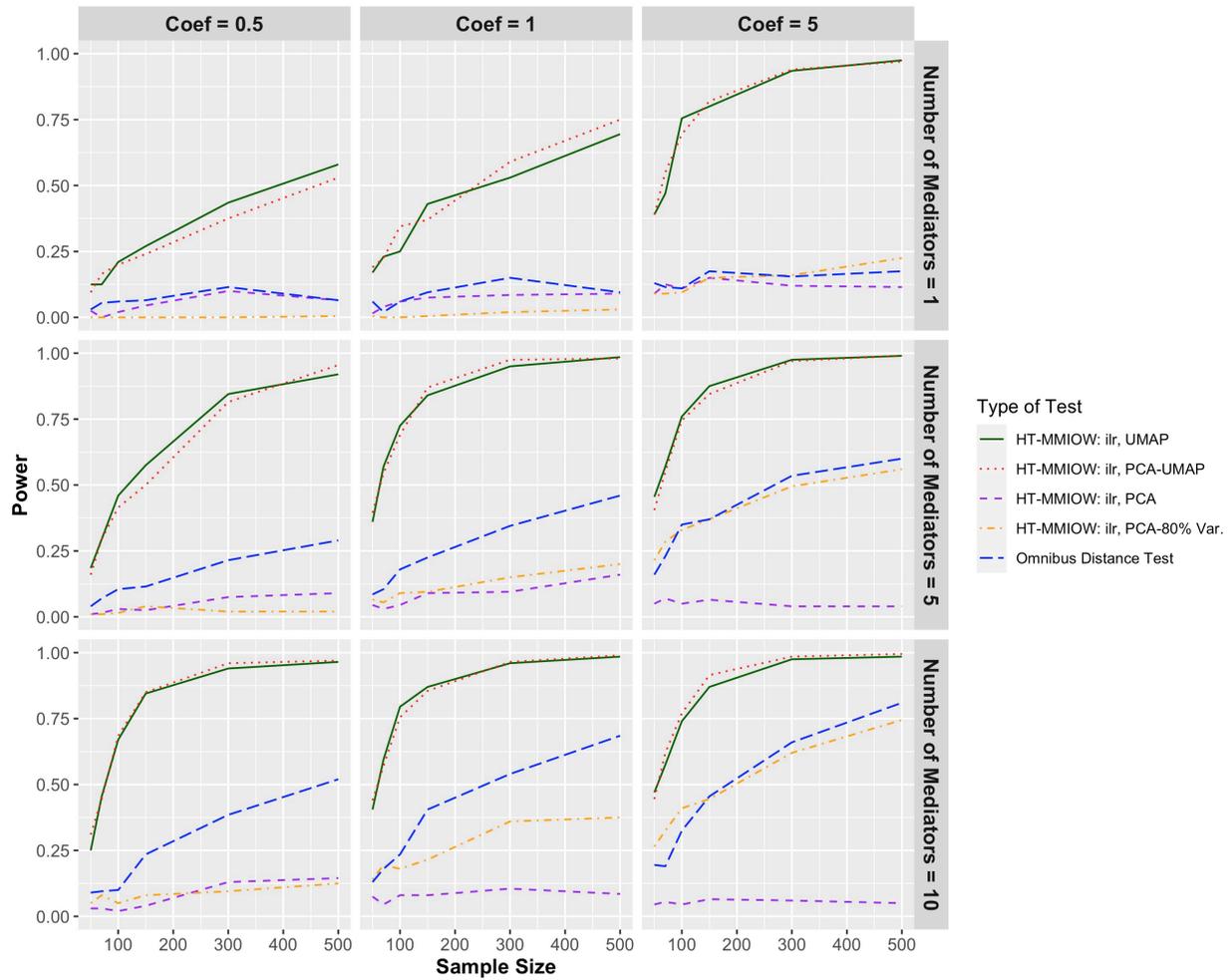

Figure 4. Power calculations for varying sample size, effect size, and number of true mediators for dichotomous outcomes and $p = 2n$. Results are based on 200 simulations. Each line represents the hypothesis test procedures evaluated.

Supplementary Material

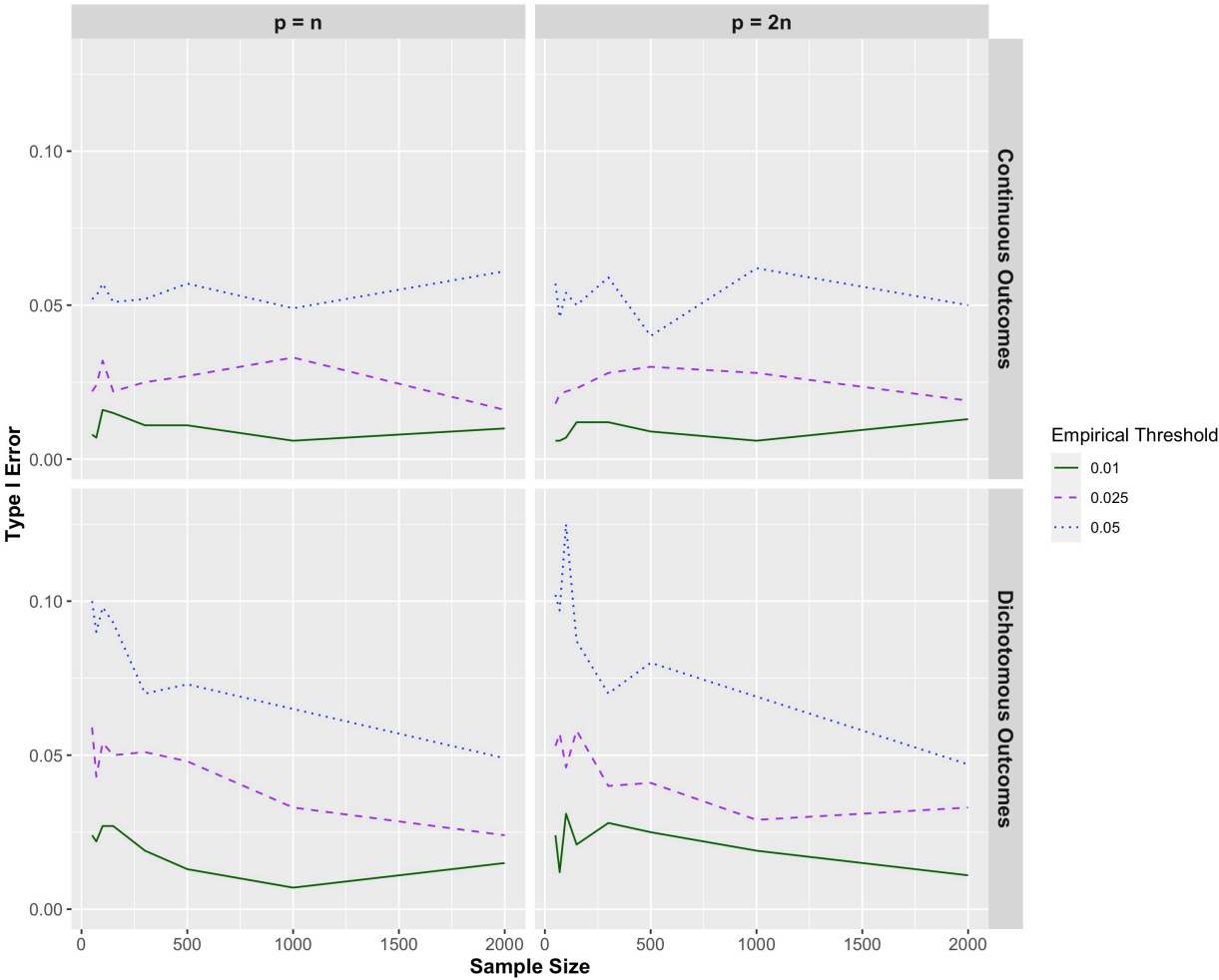

Supplementary Figure 1. Type I error calculations for varying sample size, number of taxa, and type of outcome. Results are based on 1000 simulations. Each line represents type I error at varying empirical thresholds.

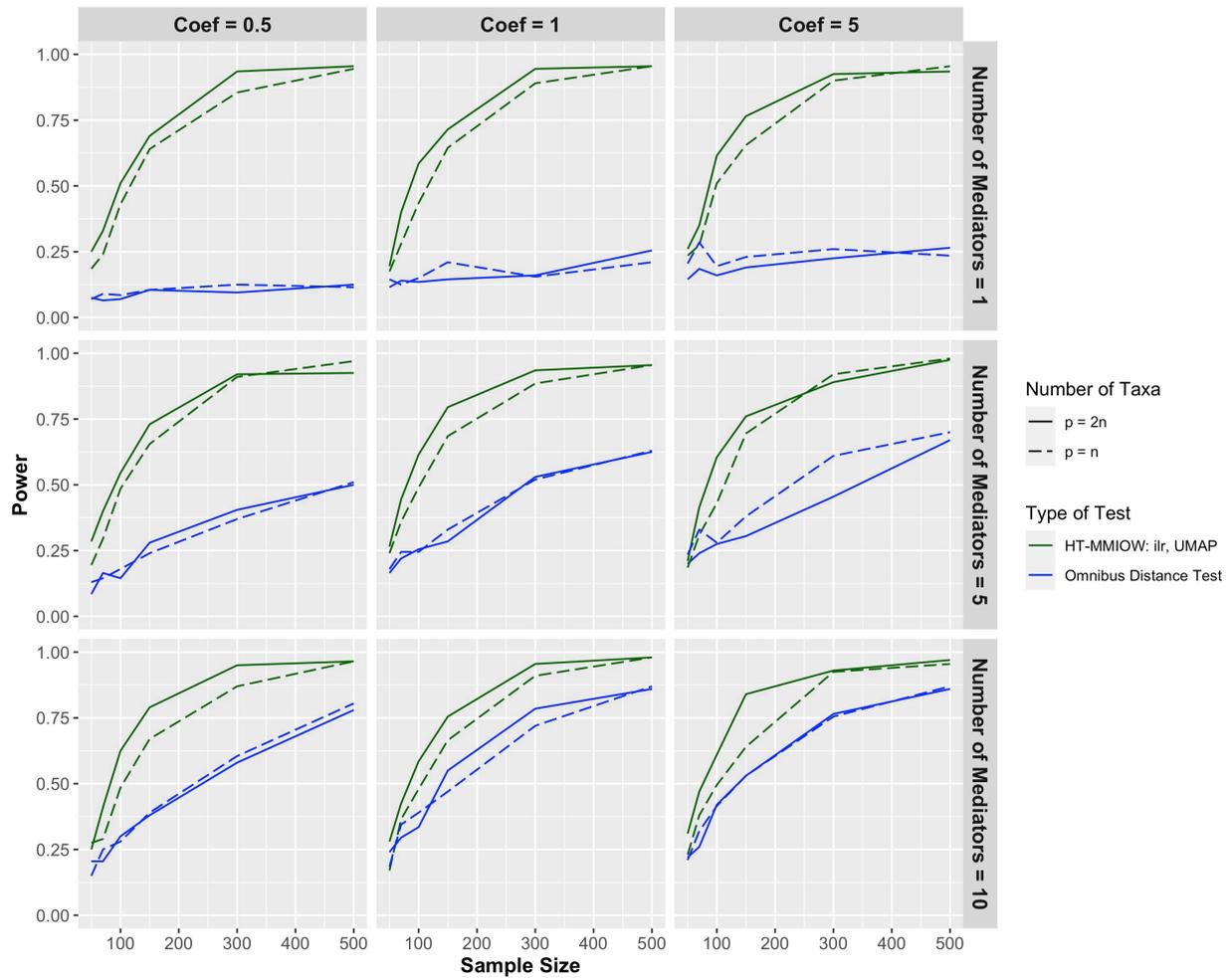

Supplementary Figure 2. Power calculations for HT-MMIOW with a 0.01 threshold and the omnibus distance test for continuous outcomes, with varying sample size, effect size, and number of true mediators and number of taxa. Results are based on 200 simulations. Solid lines indicate $p = 2n$, and dashed lines indicate $p = n$. Green lines represent HT-MMIOW with a 0.01 empirical threshold, and blue lines represent the omnibus distance test.

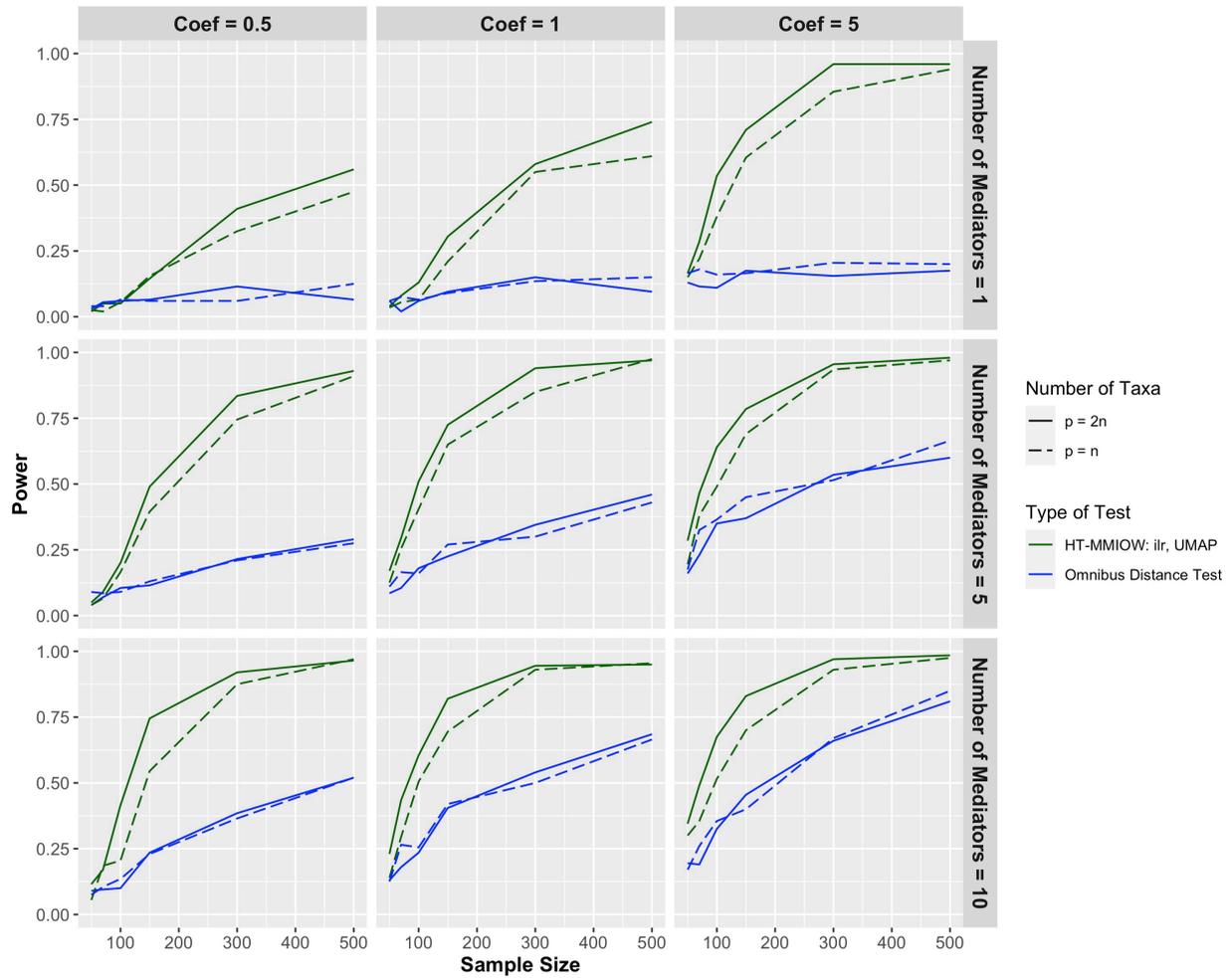

Supplementary Figure 3. Power calculations for HT-MMIOW with a 0.01 threshold and the omnibus distance test for dichotomous outcomes, with varying sample size, effect size, and number of true mediators and number of taxa. Results are based on 200 simulations. Solid lines indicate $p = 2n$, and dashed lines indicate $p = n$. Green lines represent HT-MMIOW with a 0.01 empirical threshold, and blue lines represent the omnibus distance test.